\documentclass[aps,floatfix,nofootinbib]{revtex4}

\usepackage{graphicx}
\usepackage{epic}
\usepackage{eepic}
\usepackage{latexsym}

\newcommand{\eq}[1]{(\ref{#1})}
\newcommand{\be}{\begin{equation}}
\newcommand{\ee}{\end{equation}}
\newcommand{\bea}{\begin{eqnarray}}
\newcommand{\eea}{\end{eqnarray}}

\newcommand{\vs}[1]{\vspace{#1 mm}}
\newcommand{\hs}[1]{\hspace{#1 mm}}

\newcommand{\vacr}{|0\hs{-1}>}
\newcommand{\vacl}{<\hs{-1}0|}

\def\C{\Gamma}
\def\d{\delta}
\def\D{\Delta}

\def\fr{\frac}

\def\l{\lambda}

\def\m{\mu}

\def\del{\partial}

\let\bm=\bibitem

\begin{document}
\large

\title{\Large In-in formalism, pseudo-instantons and rethinking quantum cosmology}

\author{Ali Kaya}
\email[ On sabbatical leave of absence from Bo\~{g}azi\c{c}i University. \\]{ali.kaya@boun.edu.tr}
\affiliation{\large Department of Physics, McGill University, Montreal, QC, H3A 2T8, Canada\\
and\\
Bo\~{g}azi\c{c}i University, Department of Physics, 34342, Bebek, \.{I}stanbul, Turkey\vs{15}}

\begin{abstract}
\large
Unlike flat space quantum field theories that focus on scattering amplitudes, the main observables in quantum cosmology are correlation functions. The systematic way of calculating correlators is called {\it in-in} formalism, which requires only a single asymptotic region, i.e. past infinity. The rules in perturbation theory and the path integral measure are very different for in-in and in-out formalisms, and thus the results which are standard in one approach may not necessarily hold in the other one. We show that stationary phase approximation works completely different for a scalar in-in path integral. Hence, in a cosmological background there are solutions, pseudo-instantons, that allow tunneling between locally stable vacua even in infinite volume, which is counterintuitive from an in-out perspective. We argue that various familiar notions of in-out formalism must be reexamined in the in-in formalism, which might have important consequences for quantum cosmology.  
\end{abstract}

\maketitle
\vs{25}

\centerline{Essay written for the Gravity Research Foundation 2013 Awards for Essays on Gravitation}

\vs{15}

\centerline{Received Honorable Mention}

\newpage

Rigorously defining quantum field theory is an important open problem. There are promising approaches like axiomatic formulation, which allows one to prove spin statistics and CPT theorems in flat space with Poincare invariance. Based on the concepts of locality and covariance, it is possible to extend the axiomatic method to curved spacetime \cite{w1}. However, these attempts are not yet complete and our understanding of interacting quantum fields relies on perturbation theory and a few additional non-perturbative tools. Especially in curved space, the results are sensitive to the {\it boundary conditions} and different conditions might yield dissimilar theories.  

Determining the physical observables is also an important issue since the theory must be formulated to give finite outcomes. In flat space, the main observables are scattering amplitudes and the formulation is based on the existence of asymptotic {\it in} and {\it out} particle states. In cosmology, however, the basic observables are correlation functions. Moreover, one can at most suppose the existence of a single asymptotic {\it in} region, i.e. past infinity. In some cases, even an asymptotic past infinity cannot be assumed and one should deal with a big-bang singularity or horizon appearing in the past. The systematic way of calculating correlation functions is called  {\it in-in} (or Schwinger-Keldysh or closed time) formalism \cite{i1,i2}, which is very different than the usual {\it in-out} formalism aiming to calculate the scattering amplitudes. 

It is well known that in-out and in-in perturbation theories involve quite distinct sets of rules. As we will see below,  distinctions also exist at the non-perturbative level. We will argue that these differences, which arise since one focuses on different physical quantities, may modify some standard features and lead to a 
essentially different description of field theories having a single asymptotic region as in cosmology. 
 
To briefly illustrate our point, let us consider a potential that has some form of global symmetry, which can be discrete like $Z_2$ or continuous like $U(1)$. It is well known that there appears no breaking of this global symmetry in quantum mechanics. Similarly the global symmetry is still preserved in {\it finite volume} quantum field theory. However, quantum field theory in infinite volume is different and the symmetry can be broken. 

The symmetry is preserved in quantum mechanics because the smallest eigenvalue solution of the corresponding Schr\"{o}dinger operator is unique (up to phase) if the potential tends to $+\infty$ at infinity. For a field theory in {\it finite volume}, instantons interpolating between degenerate vacua allow tunneling. In an Euclidean path integral, an instanton gives a nonzero transition amplitude that is approximately given by $e^{-S_*/\hbar}$, where $S_*$ is the value of the action for the instanton. Crucially, $S_*$ is proportional to the volume of the space and the degenerate vacua are superposed by tunneling to give a  unique vacuum state of the theory as in quantum mechanics. In infinite volume the amplitude vanishes due to the volume dependence of $S_*$ and one may construct  disjoint Hilbert spaces around each vacuum  state implying symmetry breaking (for a review of these issues see e.g.  \cite{wit}). 

\begin{figure}
\centerline{\includegraphics[width=8cm]{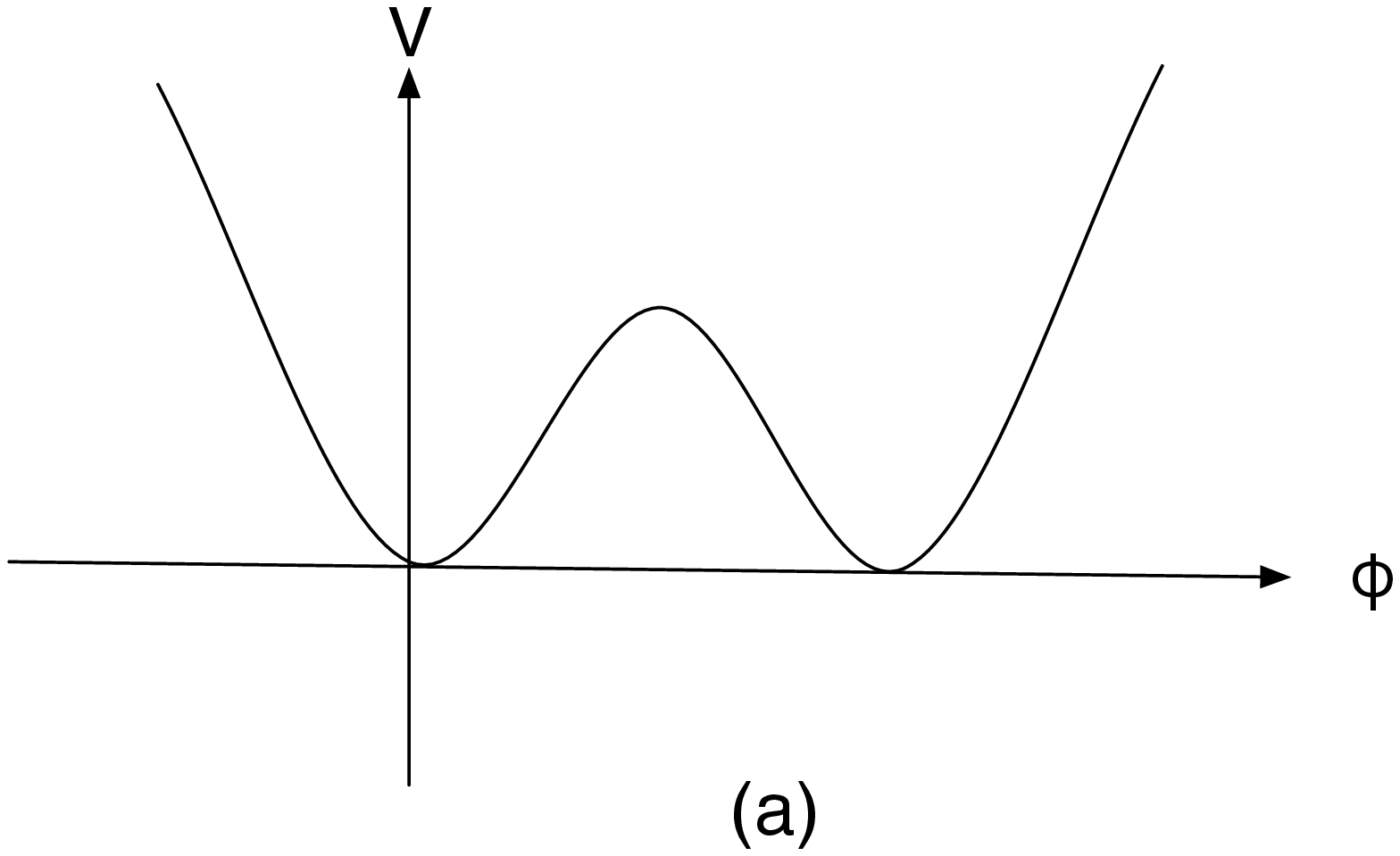}\hs{10}\includegraphics[width=8cm]{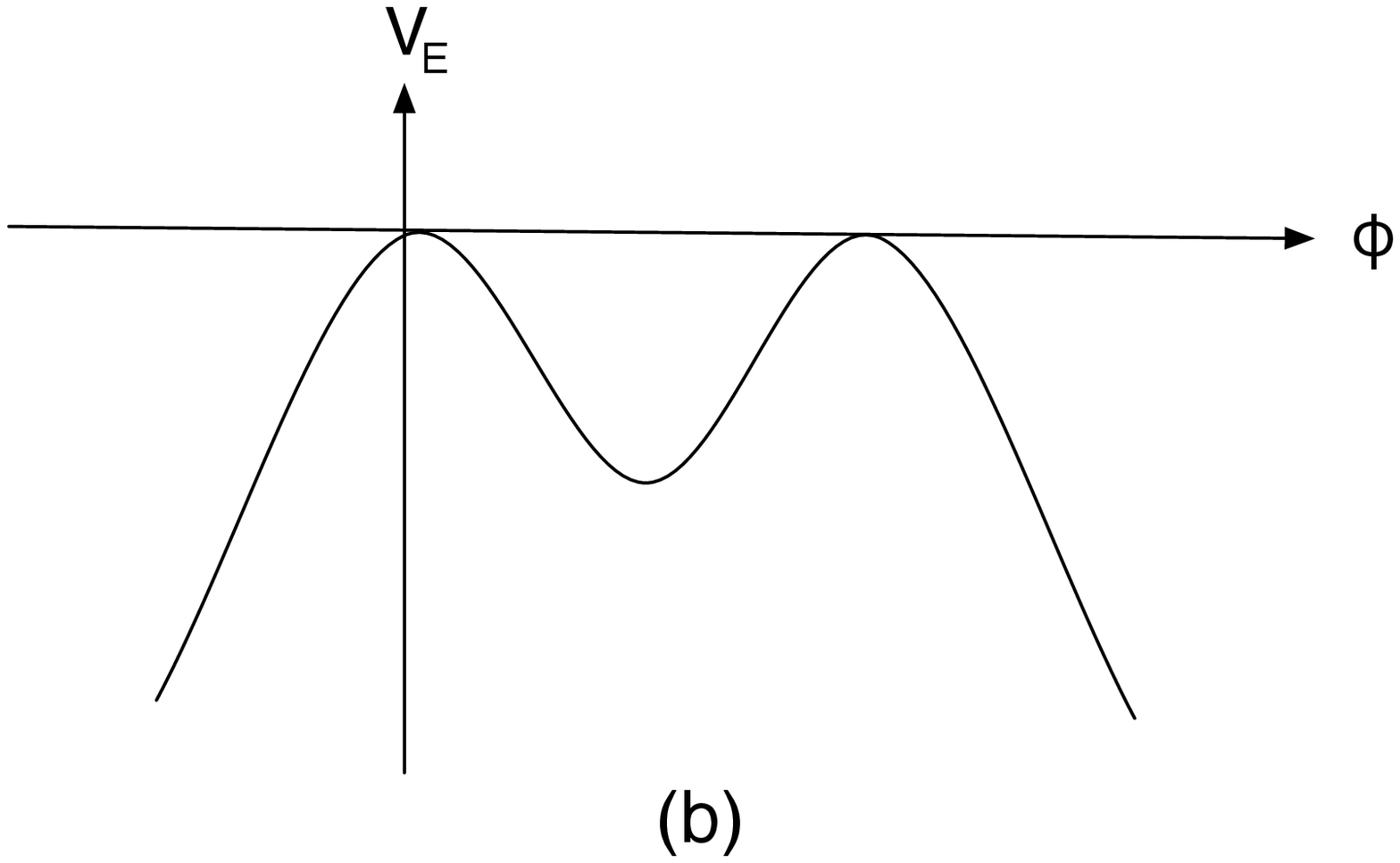}}
\caption{(a) A scalar potential with $Z_2$ symmetry having two degenerate vacua. (b) In flat space the Euclidean formulation gives an inverted potential and there exist instanton solutions interpolating between the two vacua. In finite volume,  instantons forbid symmetry breaking by tunneling.} 
\label{fig1}
\end{figure}

The above field theoretical considerations inherently uses in-out reasoning. For example, the  amplitude which is calculated by the Euclidean path integral corresponds to the transition from an {\it in} ground state at past infinity to another {\it out} ground state at future infinity. Similarly, the Hilbert spaces constructed around each vacuum state are based on the Fock space of asymptotic  particle representations. As we will discuss below, if one considers a scalar with more than one locally stable vacua in a cosmological spacetime that might have infinite spatial volume, the in-in path integral turns out to have classical stationary phase configurations interpolating between the vacua. These solutions, which we call pseudo-instantons, meaningfully contribute to the in-in path integral and like tunneling they do not allow forming disjoint Hilbert spaces around each vacuum. Thus, they are very similar to the above mentioned Euclidean instantons of finite volume quantum field theories that forbid symmetry breaking. This suggests that, as long as tunneling is considered, the field theory defined in an infinite volume cosmological background with a single asymptotic region can be similar to a finite volume in-out field theory rather than to an infinite volume one.  

Of course,  one can still use in-in formalism for the calculation of correlation functions even when an asymptotic out region exists as in flat space. In that case, a clear distinction must be made weather transition amplitudes (in-out) or correlation functions (in-in) are physical observables. Phrasing roughly, the ``physics" of correlation functions can be quite different than the ``physics'' of transition amplitudes.
 
To quantify our considerations let us take a self interacting real scalar field in a cosmological spacetime with the metric
\be
ds^2=-dt^2+a(t)^2(dx^2+dy^2+dz^2).\label{met}
\ee
The scalar action is the standard one
\be\label{a}
S[\phi,J]=-\fr12\int d^4x\sqrt{-g}\left[\nabla_\m\phi\nabla^\m\phi+V(\phi)\right]-i\int J\phi,
\ee
where a possible coupling to an external source $J$ is indicated. To avoid technical complications, we assume that the geometry nicely asymptotes as $t\to-\infty$ and there is a well defined ground state (like the Bunch-Davies vacuum) at $t=-\infty$ denoted byÊ $\vacr$. Consider the generating functional 
\be\label{g}
Z[J^+,J^-]=\int D\phi  \vacl  \phi,t_*\hs{-1}>_{J^-}<\hs{-1}\phi,t_* \vacr_{J^+},
\ee
where $J^\pm$ denote two independent sources, $|\phi,t_*\hs{-1}>$ is the complete field space basis vectors at time $t_*$ and $D\phi$ is the spatial path integral measure at $t_*$ such that $I=\int D\phi  |\phi,t_*\hs{-1}><\hs{-1}\phi,t_*|$. The transition amplitudes in \eq{g} can be expressed by path integrals so that   
\be\label{gb}
Z[J^+,J^-]=\int D\phi  \int \prod_{-\infty}^{t_*}  {\cal D} \phi^+ {\cal D}\phi^-e^{iS[\phi^+,J^+]-iS[\phi^-,J^-]} \Psi_0[\phi^+(-\infty)]\Psi_0^*[\phi^-(-\infty)],
\ee
where $\Psi_0[\phi^\pm(-\infty)]$ are vacuum-wave functionals corresponding to the inner products $<\hs{-1}\phi^\pm(-\infty) \vacr$. It is clear that differentiating $Z[J^+,J^-]$ with respect to $J^\pm$ and setting $J^\pm=0$ give various (e.g. time ordered or  anti-time ordered) correlation functions provided that one imposes $J^\pm(t_*)=0$, which is demanded by the identity operator in \eq{g} (thus $t_*$ must be chosen to be larger than any time variable in the problem). 

The path integral \eq{gb} is over all field configurations which extends from $t=-\infty$ to $t=t_*$ (the plus branch $\phi^+$) and then back, from $t=t_*$ to $t=-\infty$ (the minus branch $\phi^-$) obeying $\phi^+(t_*)=\phi^-(t_*)$. Performing the boundary path integral $D\phi$ in \eq{gb} in the discretized approximation, one can also observe that the partial time derivatives of fields must also match: $\dot{\phi}^+(t_*)=\dot{\phi}^-(t_*)$ \cite{a1}. Thus the integral is over all paths obeying
\be\label{eb} 
\phi^+(t_*)=\phi^-(t_*),\hs{7}\dot{\phi}^+(t_*)=\dot{\phi}^-(t_*).
\ee
Using \eq{eb} one may see that the path integral becomes independent of $t_*$, i.e $\del_{t_*}Z[J^+,J^-]=0$. 

The scalar in-in path integral \eq{gb} is very different compared to an analogous in-out path integral. In the former case of our interest, there is only a single asymptotic region at $t=-\infty$ for which additional boundary conditions must be imposed. The formalism uniquely fixes the boundary conditions at $t_*$, which replaces the future infinity. On the other hand, the field space is doubled, i.e. one has forward and backward propagating fields, which are denoted by $\phi^+$ and $\phi^-$, respectively. Even calculating the free path integral becomes tricky in this setup, see the appendix of \cite{we1}. 

For a self interacting scalar, one can apply perturbation theory to calculate \eq{gb} and it is possible to derive  diagrammatic rules, which are similar to Feynman graphs. There are plus and minus vertex types corresponding to $V(\phi^+)$ and $V(\phi^-)$ in \eq{gb}, and four different propagators $\D^{\pm\pm}$ related to different contractions of the fields $\phi^+$ and $\phi^-$ (the propagators are not entirely independent). 

UV renormalization of a theory is expected to be independent of the spacetime curvature once the cutoff scale is chosen much larger than the curvature scale (of course IR properties are anticipated to be altered, see \cite{a2}). Specifically a flat space renormalizable theory is expected to be renormalizable in a (smooth) cosmological spacetime with metric \eq{met}.  To some extend this was verified for $\l\phi^4$ theory in curved spacetime \cite{br1,br2} where some technical complications related to normal ordering and state dependent divergencies were solved. However, the analysis of \cite{br1,br2} was entirely in the in-out thinking where one defines an S-matrix, considers scattering amplitudes and introduces a single causal propagator. Obviously, the renormalization theory must be reexamined for the in-in formalism where one demands the finiteness of the correlation functions. Even in flat space it is not evident that the divergence of a graph with various $+$ and $-$ assignments can be canceled out by modifying the terms in the original action. The emergence of $\pm$ mixings in the counter terms is an intriguing possibility and to our knowledge this has not been checked out yet. In any case, even in the absence of these mixings, the UV structure of the theory like the cutoff dependence of the bare parameters, the beta-functions etc. are supposedly modified. Note that we are considering renormalization due to self interactions that is different than the stress energy momentum tensor renormalization as it is discussed, for example in \cite{bd}, which can be achieved by adiabatic or point splitting regularizations in the free theory. 

It is possible to illustrate the above comments in a concrete way by introducing the in-in quantum effective action  
\be
\C[\phi^+,\phi^-]=W[J^+,J^-]-\int \left(J^+\phi^+ - J^-\phi^-\right),
\ee
where as usual $W=-i\ln Z$ and $\d W/\d J^\pm=\pm\phi^\pm$. Like in the in-out case the quantum effective action has the property that the tree level diagrams of it give the exact 1PI in-in Green functions of the original theory. It is easy to see that by definition $\C[\phi,\phi]=0$, thus it is not possible to write down an effective action encoding the quantum dynamics of our original field. Instead one can show that the vacuum expectation value $\phi=\vacl \phi\vacr$ obeys 
\be\label{fe} 
\left.\fr{\d\C[\phi^+,\phi^-]}{\d\phi^+}\right|_{\phi^\pm=\phi}=0,
\ee
which now plays the role of the in-in effective equations of motion modifying classical dynamics \cite{e1}. It is precisely these field equations that should be considered in studying, for example,  the backreaction effects on the geometry and they are very different than the equations of in-out formalism. It is by no means guaranteed that an effective action can be written down whose variation would yield \eq{fe} (indeed for a general $\C[\phi^+,\phi^-]$,  \eq{fe} shows the opposite must be true). Moreover, when the background is time dependent as in \eq{met}, the parameters like the bare mass become explicitly time dependent, which demands {\it time varying}  renormalization schemes.  All these are new aspects of the in-in theory, which must be carefully analyzed.  

We should mention that most of these observations are already known and studied in some detail in many papers. The quantum effective field equations are analyzed to understand global symmetry breaking and Goldstone's theorem in an expanding universe (see e.g. \cite{sb1,sb2,sb3,sb4,sb5,sb6}). Non-perturbative techniques like the stochastic method (see e.g. \cite{s1,s2,s3}) and the dynamical renormalization group approach (see e.g. \cite{rg1,rg2}) are also utilized for the in-in theory. However, a systematic study has not yet been done to address all the technical issues like how the full renormalization theory works for the in-in perturbation theory. The existence of the  conflicting claims in the literature on some issues like de Sitter stability/instability can be attributed to this lack of understanding. 

Semiclassical methods can also be used to investigate certain aspects of field theories and instanton solutions are known to give valuable information about the vacuum structure of a theory. In searching for instantons, one usually makes a Wick rotation to the Euclidean signature and then apply the saddle point approximation. This is not an option for a general cosmological background, but one may continue to work with Lorentz signature and then apply the stationary phase approximation instead of the saddle point one (in the context of semiclassical quantum cosmology this has been studied in a less known but interesting paper \cite{singh}). 

In using stationary phase approximation for \eq{gb}, one may ignore the source couplings $J^\pm\phi^\pm$  since they are set to zero after a certain number of differentiations, which only gives a polynomial of the field variable (as it is written in \eq{a}, the source couplings do not give oscillating terms anyway).  Therefore, ignoring the vacuum wave-functionals for the moment, the phase of the integrand is given by the classical action as $S[\phi^+]-S[\phi^-]$ and it must be stationary with respect to the variations of $\phi^+$ and $\phi^-$. Since the path integral is over all fields  obeying \eq{eb},  variations must respect these conditions.

A stationary phase in this setup is given by $\Phi_{cl}=(\phi^+_{cl},\phi^-_{cl})$, i.e. it is a combination of $+$ and $-$ branches, which are not necessarily equal to each other. Varying the phase for each variable gives 
\be
\fr{\d}{\d\phi^\pm}\left(S[\phi^+]-S[\phi^-]\right)_{\Phi_{cl}}=0.
\ee
The surface terms at $t_*$ arising in this variation can be seen to vanish, thanks to the conditions \eq{eb}, and we assume that suitable boundary conditions are imposed at $t=-\infty$ so that no additional surface terms appear at $t=-\infty$. Therefore, $\phi^\pm_{cl}$ obey the classical equations of motion. Eq. \eq{eb} shows that they also have the same initial data and thus $\phi^+_{cl}=\phi^-_{cl}\equiv\phi_{cl}$. 

Till now in this discussion we have ignored the vacuum wave-functionals in \eq{gb}, which affect the functional  integrals over $\phi^\pm(-\infty)$. The vacuum wave-functionals are expected to be oscillating functionals localized around some classical vacuum configuration (for example in the free theory they are given by Gaussians localized around $\phi=0$). For $\phi^\pm(-\infty)$ functional integrals not to be negligible due to rapid oscillations, one must have
\be\label{bc}
\phi_{cl}(t)\to \phi_{vac}\hs{5}\textrm{as}\hs{5} t\to-\infty,
\ee
 where $\phi_{vac}$ denotes the vacuum value of the field. In summary,  the in-in configurations $\Phi_{cl}=(\phi_{cl},\phi_{cl})$ obeying \eq{bc}  are stationary phases of the in-in path integral, which we call pseudo-instantons \cite{a3}. 

Expanding the path integral around a pseudo-instanton we see that the zeroth order term vanishes since $S[\phi^+_{cl}]-S[\phi^-_{cl}]=S[\phi_{cl}]-S[\phi_{cl}]=0$. Therefore, there is no need to require finiteness of the action around our classical solution. This is a curious interference effect involving the cancellation of forward and backward evolving variables.  

Summarizing, a pseudo-instanton has two key properties, which are significantly different than a usual instanton solution. First, rather than obeying  Euclidean elliptic type equation, it satisfies hyperbolic differential equation where the initial value of the field variable is fixed by the vacuum chosen. Second, there is no need to require  finiteness of  the action. Since $\phi_{cl}$ obeys classical field equations and \eq{bc} fixes its value, a pseudo-instanton can be uniquely determined by the initial velocity $\dot{\phi}(-\infty)$. One may worry about imposing initial value data at infinite past. However, this is not a problem in a globally hyperbolic spacetime having constant-$t$ submanifolds as suitable initial value surfaces. For example, in the Poincare patch of the de Sitter space, one has  $a=e^{Ht}$ and  $t=-\infty$ surface is the past horizon, which is a suitable surface for imposing the initial data (recall the Penrose diagram of de Sitter space). 

\begin{figure}
\centerline{\includegraphics[width=8cm]{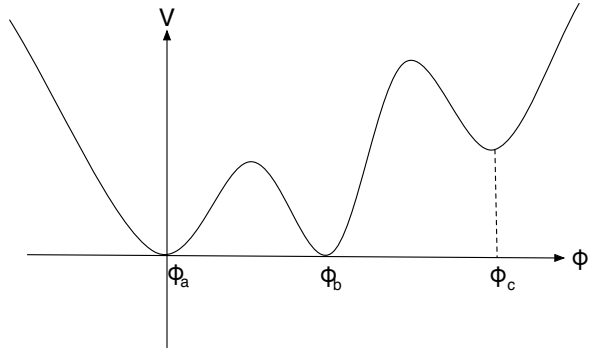}}
\caption{An example of a scalar potential supporting pseudo-instanton solutions.} 
\label{fig2}
\end{figure}

In \cite{a3}, we construct some examples of pseudo-intantons. Consider the de Sitter space that has $a=e^{Ht}$ and a potential which has different locally stable vacua, as shown in Fig. \ref{fig2}. Assume that the system is released around $\phi_a=0$ in the infinite past, i.e. $\phi=0$ vacuum is chosen.  A homogenous pseudo-instanton obeys 
\be\label{pi} 
\ddot{\phi}_{cl}+3H\dot{\phi}_{cl}+\fr{\del V}{\del \phi_{cl}}=0.
\ee
To satisfy \eq{bc} one can construct the following limiting solution defined by the initial data 
\be\label{sol} 
t_0\to-\infty:\hs{5}\phi_{cl}(t_0)=0,\hs{5}\dot{\phi}_{cl}(t_0)=v_0,
\ee
where the initial field velocity $v_0$ is not fixed by the stationary phase approximation. In \cite{a1}, such solutions are proved to be included in the path integral. For small $v_0$, $\phi_{cl}$ performs damped oscillations and settles down at $\phi=0$. However, for $v_0$ larger than a critical value, the field jumps over the first hill and it is damped around $\phi_b$. For larger $v_0$, the field can reach the pit around $\phi_c$ and performs damped oscillations around it. In the limit $t_0\to-\infty$ all these damped oscillations are pushed to infinity, which gives three static solutions $\phi=\phi_a=0$, $\phi=\phi_b$ and $\phi=\phi_c$ as pseudo-instantons. 

It is possible to expand the generating functional \eq{gb} around each pseudo-instanton implying that it can be written as $Z=Z_a+Z_b+Z_c$. For example, the scalar expectation value becomes $<\hs{-1}\phi\hs{-1}>=\phi_a+\phi_b+\phi_c$, leaving a cosmological imprint of the perturbatively inaccessible vacua. Note that the initial velocity $v_0$ is actually the zero mode parametrizing different pseudo-instanton solutions. However, as $t_0\to-\infty$ the zero mode disappears and one ends up with  three discrete solutions. 

From this discussion one may think that constructing pseudo-instanton solutions is an easy task since one should only choose the initial velocity, which is completely arbitrary. However, as discussed in \cite{a3}, this is not true for various cases in flat space since one should end up with a nice solution in the limit $t_0\to-\infty$ in \eq{sol}, which is highly restricting. 

If one believes that the path integral \eq{gb} is equivalent to the operator formalism, then the above solutions do not allow to construct disjoint Hilbert spaces around the vacua $\phi_a$, $\phi_b$ and $\phi_c$, similar to the finite volume field theory. Therefore one may claim that the quantum dynamics forbids breaking of a global symmetry, like $Z_2$ of Fig. \ref{fig1}, in cosmology. Although it is not usually underlined, the operator formalism also needs boundary conditions and the existence of a single asymptotic region seems to affect the construction of the Hilbert space. 
 
The existence of pseudo-instantons might affect the formation of topological defects in the early universe. In a typical scenario,  the effective potential develops two degenerate minima as the temperature drops below a critical value and the scalar is assumed to fall into one or the other vacuum randomly, which is expected to differ in causally separated regions in space, causing domain walls to form. But if the quantum mechanical description is still valid during this phase transition despite the presence of  thermal fluctuations  and the ``environment", then the scalar can be placed in a {\it superposed state} corresponding to the {\it true vacuum}. In that case, the expectation value of the scalar needs not to vary in space i.e. pseudo-instantons might prevent the formation of domain walls by preserving the symmetry. 

It is not difficult to generalize the above considerations to include the gravitational degrees of freedom. Although there are some technical complications due to diffeomorphism invariance, the general picture remains the same. The in-in path integral is over all doubled metric and scalar field variables $(g_{\mu\nu}^\pm,\phi^\pm)$ where the induced metric and the scalar field must match at $t=t_*$ surface for $\pm$ fields.
This time,  the spatial boundary path integral implies that the second fundamental form and the time derivative of the scalar should also agree for $\pm$ fields. For the phase of the integrand to be  stationary, $(g_{\mu\nu}^\pm,\phi^\pm)$ should obey the classical equations of motion both for $\pm$ branches and the boundary conditions at $t=t_*$ supply the same initial data showing that the $\pm$ fields are equal to the same classical configuration $(g^{cl}_{\mu\nu},\phi_{cl})$. The only remaining ingredient is to impose suitable initial conditions, possibly at $t=-\infty$ or at some finite time. In principle, the no boundary proposal of \cite{hh} can be used to determine the vacuum wave-functionals in \eq{gb}. To find the gravitational pseudo-instantons one only needs to know around which classical configuration the wave-function is localized. However, one should not attempt to calculate the wave-function of the universe today by using the no boundary proposal since the in-in path integral measure is quite different than the Euclidean measure of \cite{hh}.  

As pointed out in the first paragraph, we are still lacking a rigorous definition of quantum field theory. Generally, an information that is intended to be extracted from a path integral (or operator formalism) critically depends on some technical assumptions and the boundary conditions. There are well established and standard results, but these are mainly derived using in-out approach. In cosmology, however, the absence of an asymptotic out region forces one to reconsider the standard results and we have seen that some surprising differences can potentially alter the mainstream tenets. Although the in-in formalism does not offer a new formulation  at a fundamental level, the doubling of the field variables and a different integration measure are important deviations from the standard in-out formulation, which guides one to rethink about  some aspects of quantum cosmology.  

\
\

{\it Acknowledgements:} I would like to thank the colleagues in the theoretical high energy physics group at McGill University and especially Robert Brandenberger for their hospitality. This work is supported by T\"{U}B\.{I}TAK B\.{I}DEB-2219 grant. 

\
\

\end{document}